\documentclass[aps,prl,twocolumn,superscriptaddress]{revtex4}
\usepackage{graphicx}

\begin{document}

\title{Atom-Photon Entanglement}

\author{J\"urgen Volz}
\email{juergen.volz@physik.uni-muenchen.de}
\affiliation{Department f\"ur Physik, Ludwig-Maximilians-Universit\"at M\"unchen,
 D-80799 M\"unchen, Germany}

\author{Markus Weber}
\affiliation{Department f\"ur Physik, Ludwig-Maximilians-Universit\"at M\"unchen,
 D-80799 M\"unchen, Germany}

\author{Daniel Schlenk}
\affiliation{Department f\"ur Physik, Ludwig-Maximilians-Universit\"at M\"unchen,
 D-80799 M\"unchen, Germany}
\author{Wenjamin Rosenfeld}
\affiliation{Department f\"ur Physik, Ludwig-Maximilians-Universit\"at M\"unchen,
 D-80799 M\"unchen, Germany}
\author{Johannes Vrana}

\affiliation{Department f\"ur Physik, Ludwig-Maximilians-Universit\"at M\"unchen,
 D-80799 M\"unchen, Germany}
\author{Karen Saucke}
\affiliation{Department f\"ur Physik, Ludwig-Maximilians-Universit\"at M\"unchen,
 D-80799 M\"unchen, Germany}
\author{Christian Kurtsiefer}
\affiliation{Department of Physics, National University of Singapore, Singapore}
\author{Harald Weinfurter}
\affiliation{Department f\"ur Physik, Ludwig-Maximilians-Universit\"at M\"unchen,
 D-80799 M\"unchen, Germany}
 \affiliation{Max-Planck-Institut f\"ur Quantenoptik, 85748 Garching, Germany}

\date{\today}

\begin{abstract}
We report the observation of entanglement between a single trapped
atom and a single photon at a wavelength suitable for low-loss
communication over large distances, thereby achieving a crucial step towards long range quantum networks. To verify the entanglement we
introduce a single atom state analysis. This technique is used for full state
tomography of the atom-photon qubit-pair. The detection
efficiency and the entanglement fidelity are high enough to allow in a next step the generation
of entangled atoms at large distances, ready for a final
loophole-free Bell experiment.
\end{abstract}



\maketitle

Entanglement is a key element for quantum communication and
information applications \cite{Nielsen}. Demonstrations
of quantum computers with ions in linear chains nowadays almost
routinely create deterministically any desired entangled state
with up to four ions \cite{Roos04} and the
currently largest quantum processor consisting of some tens of
(not yet distinguishable) qubits in a so called cluster state was
implemented with neutral atoms in an optical
lattice \cite{Greiner02}. For future applications
like quantum networks or the quantum repeater \cite{Briegel98} it
is mandatory to achieve entanglement also between separated quantum
processors. For this purpose, entanglement between different
quantum objects like atoms and photons -- recently demonstrated for ions and photons \cite{Blinov04} -- forms the interface between atomic quantum memories and photonic quantum
communication channels \cite{Matsukevich04}, finally allowing the distribution of
quantum information over arbitrary distances.

Atom-photon entanglement is not only crucial for the many application of long range quantum communication, but is also the key element to give the final
answer to Einstein's question on the real properties of nature \cite{Saucke02}.
Together with Podolsky and Rosen he pointed out the
inconsistencies between quantum mechanics and their ideal of a
local and deterministic description of nature \cite{Einstein35}.
They implied that parameters of a physical system (local
hidden variables, LHV), which might not -- yet -- be known to us,
could solve the problem. Until now, the results of many
experiments based on Bell's inequality \cite{Bell64} indicate that
hidden variable theories would result in incorrect predictions and
thus are not a valid description of nature \cite{Freedman72,Clauser78,Aspect82}.
But all these tests are subject to
loopholes \cite{Clauser78,Pearle70} and none so
far could definitely outrule all alternative concepts.

Here we describe the observation of entanglement between the
polarization of a single photon and the internal state of a single
neutral atom stored in an optical dipole trap. For this purpose we
introduce a new state-analysis method enabling full state
tomography of the atomic qubit. This now allows for the first time
the direct analysis of the entangled atom-photon state formed
during the spontaneous emission process.  Moreover, we can show
that the results achieved indeed suffice to test Einstein's
objections.

\begin{figure}
\includegraphics[width=8cm]{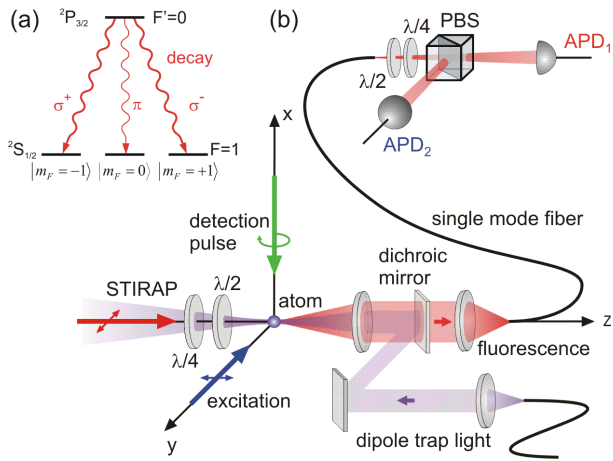}
\caption{(Color online) (a) Preparation of atom-photon entanglement in $^{87}$Rb. The excited hyperfine level with $F'=0$ can decay to three possible ground states with the magnetic quantum numbers $m_F=-1$, 0 or 1, by spontaneously emitting a $\sigma^+$, $\pi$ or $\sigma^-$ polarized photon, respectively. If the light is collected along the quantization axis, $\pi$-polarized photons are suppressed. Thus, an effective $\Lambda$ configuration is obtained which allows the preparation of a maximally entangled state between the photon polarization and the orientation of the atomic magnetic moment.
(b) Scheme of the experimental setup. The dipole trap light ($\lambda=856$ nm, $P=30$ mW) is focused by a microscope objective (NA$=$0.38) to a waist of 3.5 $\mu$m. The photon from the spontaneous decay is collected with the same objective, separated from the trapping beam by a dichroic mirror, and coupled into a single mode optical fiber guiding it to the polarization analyzer. The analyzer consists of a rotable half and quarter wave plate, a polarizing beamsplitter and two avalanche photo diodes (APD) for single photon detection. Triggered by the detection of the photon either in APD$_1$ or APD$_2$, the atomic state is analyzed using a STIRAP light field whose polarization defines the atomic measurement basis.\label{Fig1}}
\end{figure}

Atom-photon entanglement can be prepared best by exciting an atom
to a state which ideally has two decay channels
($\Lambda$-configuration). The hyperfine structure of $^{87}$Rb
offers a good approximation to such a level scheme (Fig. \ref{Fig1}(a)).
Excited to the $^2P_{3/2}$, $F'=0$ hyperfine level, the atom can
spontaneously decay into the three magnetic sublevels $|m_F=0,\pm1\rangle$ of the $^2S_{1/2}$ hyperfine level by emitting a photon at a wavelength of 780 nm.
If the emitted photon is left circularly polarized ($\sigma^-$), the atom
will be in the state $|m_F=+1\rangle$, whereas we find
$|m_F=-1\rangle$ if the emitted photon is right circularly
polarized ($\sigma^+$). Since the emitted photons are collected
along the quantization axis, $\pi$-polarized light (emitted into a
different spatial mode) is not collected for symmetry reasons and
can be ignored. As long as the remaining $\sigma^\pm$ emission
processes are indistinguishable in all other degrees of freedom
one obtains a coherent superposition of the two possible decay
possibilities, i.e. the maximally entangled state
\begin{equation}
|\Psi^+\rangle=\frac{1}{\sqrt{2}}\left(|m_F=-1\rangle|\sigma^+\rangle
 +|m_F=+1\rangle|\sigma^-\rangle\right).
\end{equation}
Here, in each of the terms the first ket describes the state of
the atom, the second one the polarization of the photon. Although the
quantum mechanical phase of this superposition follows from the Clebsch-Gordan coefficients of the transitions, it is generally not fully accepted that spontaneous emission should lead to a coherent superposition like (1) or wether not only a statistical mixture of the two possible emissions is formed. The tomography of the combined atom-photon state shows that this phase is indeed well defined.

In our experiment atoms are cooled from a shallow magneto-optical
trap (MOT) into an optical dipole trap located in the center of
the MOT. For the dipole trap waist size of 3.5 $\mu$m a
collisional blockade mechanism ensures that only single atoms are
captured \cite{Schlosser02,Weber05}. Photons emitted along the
quantization axis are collected and guided via a single mode optical fiber to a single photon
polarization analyzer to determine the state of the photonic qubit
(see Fig. \ref{Fig1}(b)).

When a single atom is loaded into the trap and its 
fluorescence is registered, the sequence
entangling the atom with a photon is started by pumping it into
the $F=1$, $m_F=0$ state. Next, a 30 ns optical $\pi$-pulse excites
the atom to the $F'=0$ level from which it will decay back to
$F=1$. The emitted photon is detected with an overall efficiency
of $\eta_{ph}\approx5\times10^{-4}$. Thus, the whole excitation
and emission process has to be repeated approximately 2000 times
which, together with intermediate cooling cycles, results in an
average rate of about 0.2 s$^{-1}$ observed atom-photon couples.

\begin{figure}
\includegraphics[width=8cm]{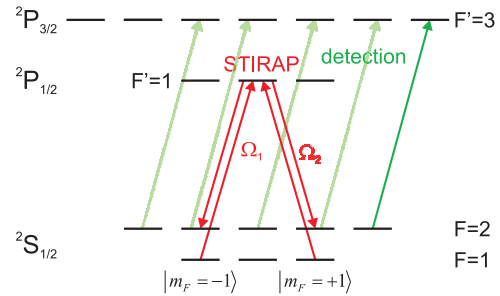}
\caption{(Color online) Experimental procedure for the atomic state detection.  To
analyze the atomic state a two-photon STIRAP-process state-selectively
transfers a superposition of the states $|m_F=-1\rangle$ and $|m_F=+1\rangle$ to the $F=2$
hyperfine level. To read out the atomic
qubit a hyperfine-level selective detection pulse is applied before
standard fluorescence detection.\label{Fig2}}
\end{figure}

Once the emitted photon is detected, the state analysis of the
atom is initiated. Standard spectroscopy techniques probing only
the populations of the states $|m_F=-1\rangle$ and
$|m_F=+1\rangle$ are not sufficient to confirm entanglement.
Instead, a projection onto general superposition states is required. We thus apply a state selective stimulated Raman adiabatic
passage (STIRAP) technique \cite{Vewinger03,Weber05} which allows
to transfer an arbitrary superposition state
$|\psi\rangle=\sin\theta|m_F=-1\rangle+e^{i\phi}\cos\theta|m_F=+1\rangle$
adiabatically to the $F=2$ ground level (Fig. \ref{Fig2}). Due to selection rules of
atomic dipole transitions the orthogonal quantum state does not
couple to the STIRAP light field $\Omega_1$ and remains in the
$F=1$ level. The angles $\theta$ and $\phi$ in this process are
defined by the relative amplitude and phase of the $\sigma^+$ and
$\sigma^-$ polarization components of the STIRAP laser $\Omega_1$,
respectively. In essence, the polarization of the STIRAP laser
defines which superposition state is transferred, thus allowing a
full tomographic analysis of the atomic state without the
necessity to perform any state manipulation on the atomic qubit.

\begin{figure}
\includegraphics[width=8cm]{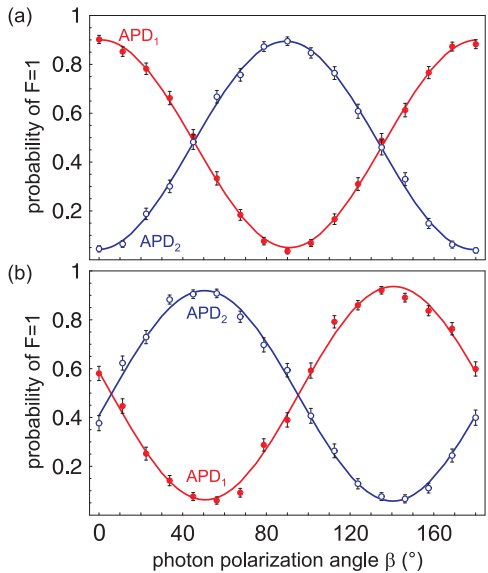}
\caption{(Color online) Probability of detecting the atom in the ground level $F=1$ (after the STIRAP pulse) conditioned on the detection of the photon in detector APD$_1$ (-$\bullet$-) or APD$_2$ (-$\circ$-) as the linear polarization of the photonic qubit is rotated by an angle $\beta$. (a) The atomic qubit is measured in $\hat\sigma_x$; (b) in $\hat\sigma_y$, whereas the photonic qubit is projected onto the states $1/\sqrt{2}(|\sigma^+\rangle\pm e^{2i\beta}|\sigma^-\rangle)$.\label{Fig3}}
\end{figure}

After the STIRAP pulse the atom is in a superposition of the
hyperfine ground levels $F=1$ and $F=2$ which now can be
distinguished by standard methods. We apply a laser pulse
(resonant to the closed transition $F=2\rightarrow F'=3$) removing
atoms in the $F=2$ level from the trap. Finally, to read out the
atomic state the cooling lasers of the MOT are switched on and
atomic fluorescence is measured for 30 ms to decide whether the
atom is still in the trap or not. Thereby, we obtain the binary
result of the projective atomic state measurement on the state
$|\psi\rangle$ and the orthogonal state $|\psi_\bot\rangle$. For the results shown in Fig.
\ref{Fig3} we repeated the experimental cycle approximately 300 times per
data point from which we obtain the probability of the atom to
remain in $F=1$ with a statistical error of $\pm 2\%$.

To verify the entanglement of the generated atom-photon state we
perform $\hat\sigma_x$ ($\theta=\pi/4$, $\phi=0$) as well as
$\hat\sigma_y$ ($\theta=\pi/4$, $\phi=\pi/2$) state analysis of the
atomic qubit for different polarization measurements of the photon
(Fig. \ref{Fig3}, $\hat\sigma_i$ are the spin-$1/2$ Pauli operators). 
Thereby, the probability of the atom to be transferred by the STIRAP pulse
sequence, or the probability to remain in the $F=1$ ground level,
respectively, is measured, conditioned on the polarization measurement outcome of the photon. Varying the photon polarization analyzer, this probability shows the expected sinusoidal dependence for both $\hat\sigma_x$ and $\hat\sigma_y$.
From the fits to the measured data we
obtain an effective visibility (peak to peak amplitude) of
$V=0.85\pm0.01$ for analysis in $\hat\sigma_x$ and $V=0.87\pm0.01$ for
analysis in $\hat\sigma_y$. This clearly proves entanglement of the
generated atom-photon state.

\begin{figure}
\includegraphics[width=8cm]{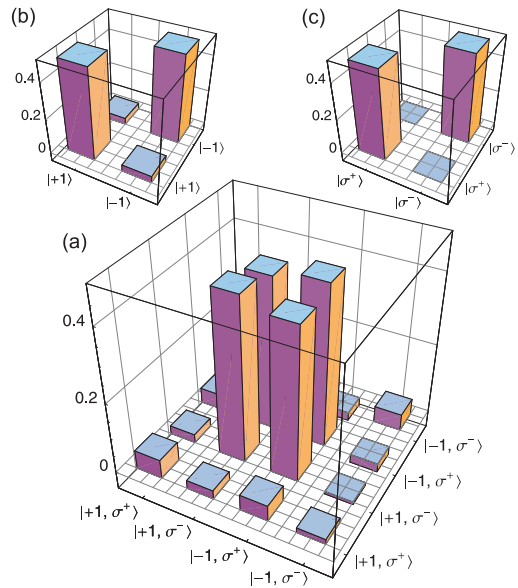}
\caption{(Color online) (a) Graphical representation of the real part of the measured density matrix of the entangled atom-photon state. The fidelity (overlap with the expected state $|\Psi^+\rangle$) from this measurement is $F=0.875\pm0.012$. Inset (b) and (c) show the single particle density matrices for the atom and photon state, respectively, indicating that the single particles when observed on their own are found in a completely mixed state.\label{Fig4}}
\end{figure}

For the determination of the full atom-photon state we 
perform two-qubit state tomography. This involves the measurement
of all combinations of the operators $\hat\sigma_x$, $\hat\sigma_y$, and
$\hat\sigma_z$ on the atom and the photon \cite{James01}. The density
matrix $\rho_{at-ph}$ determined this way clearly proves the state
to be of the form of (1) (see Fig. \ref{Fig4} (a)). The fidelity, defined as the overlap between
$|\Psi^+\rangle\langle\Psi^+|$ and $\rho_{at-ph}$, is $F=0.87\pm0.01$.
Applying the Peres-Horodecki criterion \cite{Peres96} to the combined density
matrix proves the entanglement with a negativity of $0.382$. Fig. \ref{Fig4} (b) and \ref{Fig4} (c) show the density matrices of the atomic and the
photonic state after tracing over the partner qubit. Obviously, these states are completely mixed states. This is what was observed with standard spectroscopy. However,
from our experiment it becomes clear that the resulting
atom-photon state is not a  mixture of all possible contributions but is
instead a well defined (ideally) pure, entangled state.

In view of these results let us now analyze the performance of a
possible loophole-free Bell experiment with a pair of entangled
atoms. Crucial for such a test is a highly efficient state analysis
by space-like separated observers. 
To generate entanglement between atoms at remote locations they are first entangled with a photon each. The two photons are brought together and then are subject to a Bell-state measurement, which serves to swap the entanglement to the
atoms \cite{Zukowski93}.
Starting with two entangled atom-photon pairs each in a state
with visibility $V_{at-ph}$, the visibility of the entangled
atom-atom state (after entanglement swapping) is ideally given
by $V_{at-at}=V_{at-ph}^2$ \cite{Duer99}. If we use the average
visibility observed in our experiment we thus derive an expected atom-atom
visibility of $V_{at-at}=0.74\pm0.01$. In comparison with related
experiments \cite{Pittman03, Legero04} we assume
that the Bell-state analysis, required in the entanglement swapping
process for narrowband photons and single-mode fibers, can be
performed with a fidelity of better than 0.98. Thus, the violation
of a Bell inequality, which is achieved above the threshold
visibility of $0.71$ for a CHSH-type Bell's
inequality \cite{Clauser69}, is feasible.

We emphasize, that triggered on the detection of a photon every
atomic state measurement yields a result. In this sense, the
detection efficiency (the probability to obtain a result from the
atomic state measurement) here is equal to one. In certain
cases, as e.g. the loss of the atom from the trap, the measurement
might give wrong results reducing the visibility, but one always
obtains a result. The raw data presented above of course contains
such cases, nevertheless, the visibility is high enough. Moreover, entanglement swapping enables a so called event-ready scheme \cite{Clauser78,Bell88,Zukowski93}. If measurement results are reported for every joint photon detection event, this scheme is independent of any additional
assumptions and thus is not subject to any detection related loopholes at
all.
To close at the same time the locality loophole, the atoms have to
be space-like separated with respect to the measurement time of
the atomic states. The minimum distance of the atoms is determined
by the duration of the whole measurement sequence, here mainly
given by the atomic state detection. In our experiment the
superposition of the atomic hyperfine states collapses by
scattering photons from the detection laser. After approximately
10 lifetimes ($\tau=26$ ns) of the $^2P_{3/2}$ excited state the
reduction of the initial superposition is completed with a
probability of more than 99$\%$. Together with the STIRAP-process
this yields an overall measurement time of less than 0.5$\mu$s
requiring a separation of the atoms of 150 m. 
The generation of entangled atom-photon pairs is probabilistic
with a success probability given by the total detection efficiency
$\eta_{ph}$ of the emitted photons. Taking into account
transmission losses of the photons ($T^2(75m)=0.9$, repetition rate
for that distance: $5\cdot 10^5s^{-1}$) we expect the generation of
about one entangled atom-atom pair per minute \footnote{Because the atomic state detection has to be performed only when a photon pair event was registered, the repetition rate will be significantly higher than expected from the square of the success probability of atom-photon generation.}. Then, a
loophole-free violation of e.g. a CHSH-type Bell's
inequality \cite{Clauser69} by three standard deviations, requiring
approximately 7000 atom pairs at the expected visibility of
0.74, would be feasible with a total measurement time of 12 days.

In this contribution we presented a successfull implementation of
a source of high-fidelity entangled atom-photon pairs. We introduced a single atom STIRAP state analysis which does not require additional atomic state manipulations and thus can be performed with increased fidelity. This allowed us to perform the first full state tomography of an
atom-photon system and proved that the spontaneous emission of the
atom results in the entangled state $|\Psi^+\rangle$. In the
experiment we achieved a state fidelity of $F=0.87\pm0.01$ and a mean visibility of the
atom-photon correlations of $V_{at-ph}=0.86\pm0.01$. These methods,
possibly combined with high-Q cavities to enhance the collection
efficiency \cite{Legero04,Kimble04}, form the basic elements in
future quantum information experiments for building the interface
between quantum computers and a photonic quantum communication
channel. In addition, these tools also help to find an answer to
the long standing question whether local realistic extensions
of quantum mechanics can describe nature at all. The experimental
demonstration of high-fidelity entanglement provides the most
important step towards a final, loophole-free test of Bell's
inequality.

\acknowledgements{We acknowledge stimulating discussions with T. W. H\"ansch and his group. This
work was supported by the Deutsche Forschungsgemeinschaft and the
European Commission through the EU Project QAP (IST-3-015848).}

\end{document}